\documentclass[12pt,preprint]{aastex}

\slugcomment{To appear in the Astrophysical Journal.}
\slugcomment{\bf DRAFT: \today}
\shorttitle{Intermediate Velocity Gas toward HD~14434}
\shortauthors{Knauth et al.}

\begin{document}

\title{On the Origin of the High-Ionization Intermediate-Velocity Gas
  Toward HD~14434\altaffilmark{1}}

\author{David C. Knauth\altaffilmark{2},
J. Christopher Howk\altaffilmark{2}$^,$\altaffilmark{3}, 
Kenneth R. Sembach\altaffilmark{4}, James T. Lauroesch\altaffilmark{5}, 
and David M. Meyer\altaffilmark{5}}

\altaffiltext{1}{Based on observations made with the NASA/ESA Hubble Space
Telescope, obtained from the data archive at the Space Telescope
Science Institute. STScI is operated bythe Association of Universities
for Research in Astronomy, Inc. under the NASA contract NAS 5-26555. }

\altaffiltext{2}{{\it FUSE} Science Center, Department of Physics and
  Astronomy, The Johns Hopkins University, Baltimore, MD 21218;
  dknauth@pha.jhu.edu.}

\altaffiltext{3}{Current address: Center for Astrophysics and Space
  Sciences, University of California at San Diego, C-0424, La Jolla,
  CA, 92093; howk@trafalgar.ucsd.edu.}

\altaffiltext{4}{Space Telescope Science Institute, 3700 San Martin
  Dr., Baltimore, MD 21218; sembach@stsci.edu.}

\altaffiltext{5}{Department of Physics and Astronomy, Dearborn
  Observatory, 2131 Sheridan Road, Northwestern University, Evanston,
  IL 60208; jtl@elvis.astro.nwu.edu; davemeyer@northwestern.edu.}  
{}

\vfill

\begin{abstract}
  
  We present {\it Far Ultraviolet Spectroscopic Explorer} and Space
  Telescope Imaging Spectrograph observations of high-ionization
  interstellar absorption toward HD~14434 [($l$, $b$) = 
  (135.$\!\!^{\circ}$1, -3.$\!\!^{\circ}$8); d $\sim$ 2.3 kpc], an O5.5 V 
  star in the Perseus OB1 Association.  Intermediate-velocity interstellar
  \ion{Si}{4} and \ion{C}{4} absorption is present at $V_{LSR}$
  = -67~km~s$^{-1}$, while low-ionization gas associated with the
  Perseus arm is detected at $\sim$ -50 km s$^{-1}$.  Neither
  \ion{N}{5} nor \ion{O}{6} is detected at $V_{LSR}$ = -67 km
  s$^{-1}$; although \ion{Al}{3} and \ion{Fe}{3}, tracers of
  warm ionized gas, are seen.  The high-ion column densities in the
  -67~km~s$^{-1}$ component are $\log$[$N$({\rm \ion{C}{4}})]~=~13.92
  $\pm$ 0.02 cm$^{-2}$, $\log$[$N$({\rm \ion{Si}{4}})]~=~13.34 $\pm$ 0.02
  cm$^{-2}$, $\log$[$N$({\rm \ion{N}{5}})] $\leq$ 12.65 cm$^{-2}$, and 
  $\log$[$N$({\rm \ion{O}{6}})] $\leq$ 13.73 cm$^{-2}$ (3-$\sigma$ limits).
  The observed \ion{C}{4}/\ion{Si}{4} ratio of 3.8 $\pm$ 0.3 in this
  intermediate-velocity cloud (IVC) is similar to the Galactic average
  (4.3 $\pm$ 1.9).  Our analysis of the \ion{Si}{4} and \ion{C}{4}
  line widths yields a temperature of T $\sim$ 10,450 $\pm$
  3,400~K for this component.  At this low temperature, neither
  \ion{Si}{4} nor \ion{C}{4} can be produced via collisions.  We
  investigate several photoionization models to explain the
  intermediate-velocity \ion{Si}{4} and \ion{C}{4} absorption toward
  HD~14434.  Photoionization models employing cooling of a hot ($T$ $\sim$ 
  10$^6$~K) diffuse plasma as the source of ionizing radiation reproduces the
  observed properties of the IVC toward HD~14434 quite well.  The hot plasma
  responsible for the ionizing radiation in these models may be attributed to
  hot gas contained in a supershell in or near the Perseus Arm or from a more
  generally distributed hot ionized medium.
\end{abstract}


\section{Introduction}

The canonical picture of the interstellar medium (ISM) in the Galactic
disk is one in which the gas resides in multiple ``phases'' in rough
pressure equilibrium with one another.  While the details vary (Cox \&
Smith 1974; McKee \& Ostriker 1977), all global models of the ISM in
the Milky Way include a hot component ($T\sim10^5$ to $>10^6$ K).
This hot component, motivated by the detection of Li-like oxygen (\ion{O}{6})
absorption (Rogerson et al. 1973) and X-ray emission
(Williamson et al. 1974) from the ISM some 30 years ago, is maintained
by the input of energy and matter from massive stars into the ISM through 
stellar winds and supernovae.  The precise nature of this feedback into the ISM
is not well understood, although it is thought to provide for,
among other things, the energy required to explain the quantity and
distribution of the ``high-ions'' \ion{O}{6}, \ion{N}{5}, \ion{C}{4},
and \ion{Si}{4} in the Milky Way (e.g., Savage, Sembach, \& Lu 1997;
Savage et al. 2002).

OB associations containing young, massive stars are ideal locations
for studying the hot interstellar medium, since these stars are
believed to be directly responsible for its production.  The massive
Perseus OB1 Association resides in the Perseus Arm of the Milky Way,
some 2.3 kpc from the Sun \citep{GS92}.  The {\it International
Ultraviolet Explorer} ({\it IUE}) detected high-velocity absorption in the
highly-ionized species \ion{Al}{3}, \ion{Si}{4}, and 
\ion{C}{4} toward several stars in this association (Phillips \&
Gondhalekar 1981; McLachlan \& Nandy 1985; Savage, Meade, \& Sembach
2001).  \citet{PG81} and \citet{MN85}
concluded that this highly-ionized gas was caused by a supernova
explosion within Per OB1 itself.  This region is rife with evidence
of stellar feedback on massive scales, including the presence of
\ion{H}{1} and H$\alpha$ supershells (Heiles 1979; Reynolds, Sterling,
\& Haffner 2001; Madsen, Haffner, \& Reynolds 2002), lending 
indirect support to their conclusion.  However, the physical processes
directly responsible for the production of the observed highly-ionized
gas are unconstrained.

In this work, we focus on the ionized gas observed toward
HD~14434, a member of the Perseus OB1 Association.  High-resolution
Space Telescope Imaging Spectrograph (STIS) and {\it Far Ultraviolet
  Spectroscopic Explorer} ($FUSE$) observations of interstellar
absorption associated with the Perseus Arm toward HD 14434 are
presented.  This sight line is remarkable for the presence of an
intermediate-velocity cloud (IVC), the velocity of which is
inconsistent with its origin in the quiescent gas of the Perseus Arm.
This cloud is seen strongly in the highly-ionized species
\ion{Si}{4} and \ion{C}{4}; it shows some absorption from the
moderately-ionized species \ion{Al}{3} and \ion{Fe}{3} and
none from the more highly-ionized species \ion{N}{5} and \ion{O}{6}.
While this cloud was seen with {\em IUE}, the echelle-mode STIS
observations presented here constrain its temperature to a range that
is indicative of photoionized material, and we will show that this
material is likely ionized by emission from a diffuse X-ray emitting
hot plasma.

This paper is organized as follows.  A description of the HD14434
sight line is given in \S2. The STIS and $FUSE$ observations and data
reduction techniques are described in \S3.  In \S4, the absorption
line measurements are discussed.  The nature of the highly-ionized gas
toward HD~14434 is presented in \S5.  Our conclusions are summarized
in \S6.

\section{The Sight Line toward HD~14434}

\begin{figure}\figurenum{1}\epsscale{0.8}
\begin{center}
\plotone{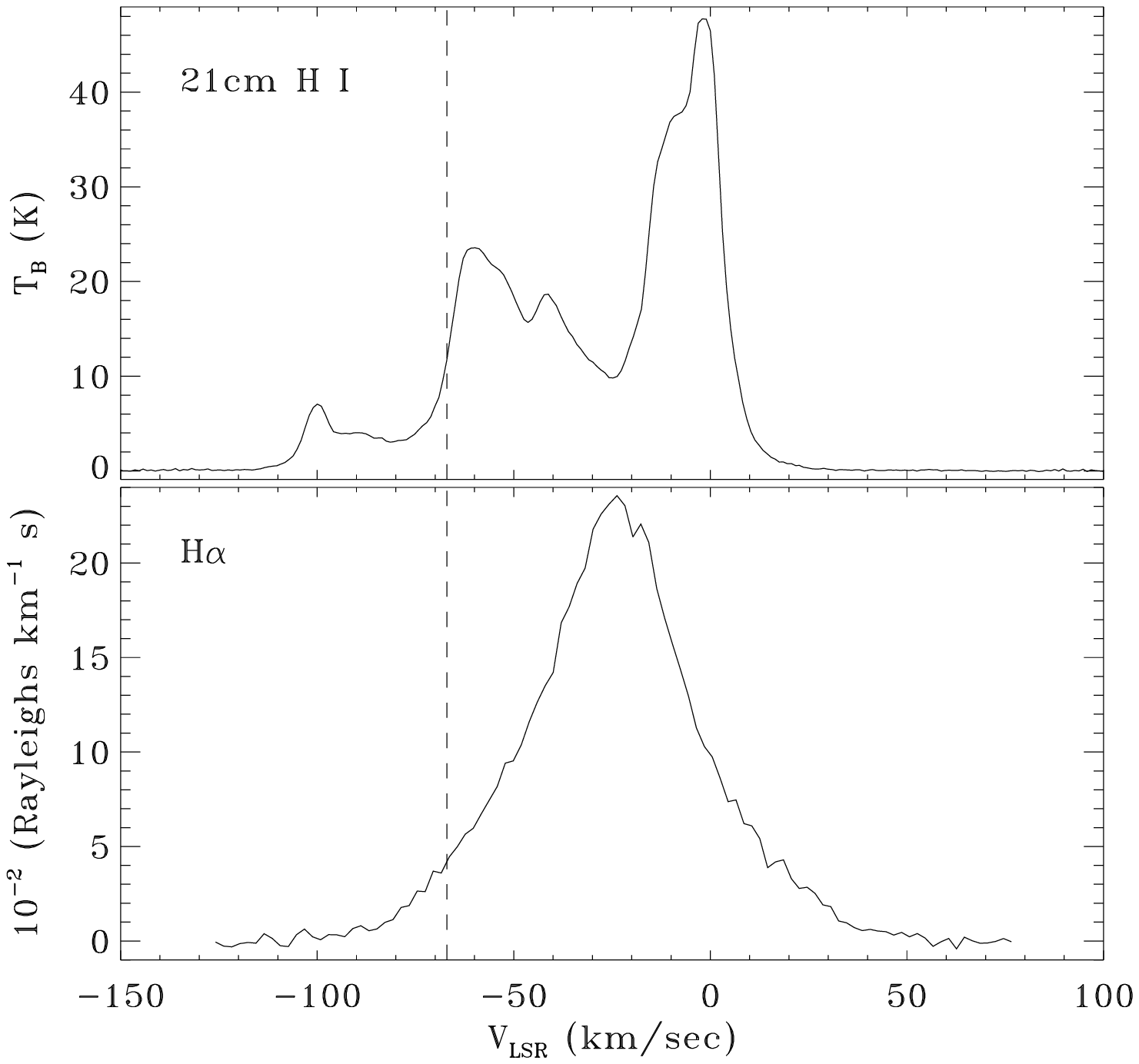}
\vspace{0.5in} 
\caption{\label{HI_WHAM} 21cm \ion{H}{1} emission \citep{HB97} and H$\alpha$ 
WHAM data \citep{haf01} toward HD~14434 are presented.  The \ion{H}{1} data 
clearly reveal two components from the Local (V$_{lsr}$ $\sim$ 0 km s$^{-1}$)
and Perseus Arms (V$_{lsr}$ $\sim$ -50 km s$^{-1}$) and a single weak 
component associated with the Outer Arm (V$_{lsr}$ = -100 km s$^{-1}$).
The H$\alpha$ data show a single broad component centered at V$_{lsr}$ = 
-25 km s$^{-1}$.  The vertical line is at V$_{lsr}$ = -67 km s$^{-1}$, the 
velocity of the IVC.  Neither \ion{H}{1} nor H$\alpha$ reveal any clues as to 
the presence of the IVC at V$_{lsr}$ = -67 km s$^{-1}$.}
\end{center}
\end{figure}

HD~14434, an O5.5 Vnfp star (Walborn 1972), is a member of the Per OB1
Association, which is a region of active star formation.  Table~\ref{star_par} 
contains the relevant stellar parameters used throughout this work.  We assume 
the distance to HD~14434 is 2.3 kpc, the distance to Per~OB1 \citep{GS92}.   
At this distance the line of sight toward HD~14434 passes through both the Local
Arm ($V_{LSR}$ $\sim$ 0 km s$^{-1}$) and the Perseus Arm ($V_{LSR}$ $\sim$ -50 
km~s$^{-1}$).  Figure~\ref{HI_WHAM} shows the 21cm \ion{H}{1} emission results
of the Ledien/Dwingeloo Survey \citep{HB97} and H$\alpha$ emission obtained by
WHAM, the Wisconsin H$\alpha$ Mapper, \citep{haf01} toward HD~14434.  The 21cm 
\ion{H}{1} data clearly reveal two components from the Local and Perseus Arms 
and a single weak component associated with the Outer Arm (V$_{lsr}$ = -100 km
s$^{-1}$).  The H$\alpha$ data show a single broad component centered at 
V$_{lsr}$ = -25 km s$^{-1}$.  At V$_{lsr}$ = -67 km s$^{-1}$ (vertical line in 
Figure~\ref{HI_WHAM}), the velocity of the IVC studied here, neither \ion{H}{1} 
nor H$\alpha$ reveal any clues as to the presence of the IVC.  

Evidence for star-ISM interaction is prevalent in this region.  In addition to 
being an active star forming region, Per~OB1 resides in close proximity to the 
double cluster $h$ and $\chi$ Per [NGC~869 at ($l$, $b$) =
(134.$\!\!^{\circ}$5, -3.$\!\!^{\circ}$5) and NGC~884 at ($l$, $b$) =
(135.$\!\!^{\circ}$1, -3.$\!\!^{\circ}$6)].  These open clusters are
believed to be related to Per OB1, with evidence for three episodes of
star formation in this localized region of the Perseus Arm (Schild
1967; Marco \& Bernabeu 2001).  High-resolution H~{\small I} observations show 
the presence of numerous wind-blown interstellar bubbles around many stars in this region
(Cappa \& Herbstmeier 2000), including HD~13022, HD~13338, HD~14442, and 
HD~14947, all members of Per~OB1.  Additionally, Heiles
(1979) has identified two Galactic supershells in the vicinity of Per
OB1: GS139-03-69 at $l$~=~139$^{\circ}$ and $b$~=~-3$^{\circ}$ with
$\Delta$$l$ = 18$^{\circ}$ and $\Delta$$b$ = 10$^{\circ}$, and one unnamed
shell 5$^{\circ}$ in diameter centered on Per OB1 at $l$~=~135$^{\circ}$ and
$b$~=~-4$^{\circ}$ (Heiles 1979).  Following Heiles's naming convention, we 
refer to this shell as GS135-04-27.  The GS139-03-69 supershell is visible in
the velocity range -87 $\le$ V$_{LSR}$ $\le$ -59 km s$^{-1}$ (Heiles 1979; Knee 
\& Brunt 2001), while the observed velocity range for GS135-04-27 is -31 $\le$ 
V$_{LSR}$ $\le$ -23 km s$^{-1}$ (Heiles 1979).  The \ion{H}{1} emission 
detected between -60 and -70 km s$^{-1}$ in the vicinity around ($l$, $b$) 
$\sim$ (135.$\!\!^{\circ}$0, $b$=-3.$\!\!^{\circ}$5) is likely associated with 
the southern rim of the GS139-03-69 (Knee 2003, private communication). 
However, the standard Galactic rotation model predicts a distance to 
GS139-03-069 of 9 kpc (Kerr \& Lynden-Bell 1986) much further away than Per~OB2.
It is important to note that an accurate distance cannot be obtained from 
velocity information alone, therefore an association between GS139-03-69 and 
our IVC cannot be ruled out.   Higher spatial resolution \ion{H}{1} and 
H$\alpha$ data toward HD~14434 and Per~OB2 are required to obtain further 
insight. 

Recent observations from the Wisconsin H$\alpha$ Mapper (WHAM) detect
a supershell-like structure in the velocity range -60 km s$^{-1}$
$\leq$ $V_{LSR}$ $\leq$ -40 km s$^{-1}$ approximately centered on Per OB1
(Madsen, Haffner, \& Reynolds 2002).  This velocity range is similar to that 
seen in intermediate and high-ion absorption (Philips \& Gondhalekar 1981; 
McLachlan \& Nandy 1985; this work).  The extended structure seen south of the
plane in the Perseus Arm may be related to the northern supershell
studied by Reynolds, Sterling, \& Haffner (2001).

The large-scale structures identified in \ion{H}{1} and H$\alpha$
emission are likely related to the vigorous star formation occuring in the
Galactic disk near this region.  In particular, the W4 \ion{H}{2} region is
found at ($l$, $b$)~=~($134\fdg7$, $0\fdg9$), and there is a large
amount of evidence that the stars in this region are influencing the
ISM on large scales (Normandeau, Taylor, \& Dewdney 1996).  Star formation
associated with W4 may be responsible for driving the vertically-extended 
structures seen in the H$\alpha$ maps of Reynolds et al. (2001) and Madsen 
et al. (2002).  While the observations summarized above are not focused on the
specific sight line toward HD~14434, it is important to note that the
region in which HD~14434 resides is strongly affected by the presence
of massive stars.  

\section{Observations and Data Reduction}
\subsection{STIS}

HD 14434 was observed on 2001 April 6, by STIS onboard the
{\it Hubble Space Telescope} ({\it HST}) as part of the guest observer
proposal ``Snapshot Survey of the Hot ISM'' (ID 8662).  The data were
acquired with the E140M grating and the 0.2$^{\prime\prime}$ x
0.2$^{\prime\prime}$ aperture for a total exposure time of 1440
seconds.  This setup provides 43 echelle orders covering the complete
wavelength range from 1150 $-$ 1725 \AA\ at a resolving power of
$R$~$\sim$~46,000 ($\Delta$$v$ = 6.5 km s$^{-1}$).  There are two
pixels per resolution element.  The data were reduced and extracted
with the CALSTIS pipeline (v2.11).  The subtraction of background and scattered 
light from these echelle data employs the algorithm of Lindler \& Bowyers 
(2000).  CALSTIS provides the appropriate Doppler correction to remove the 
effect of the spacecraft motion and places the spectra on the heliocentric 
velocity scale ($V_{helio}$), accurate to 0.8--1.6 km s$^{-1}$.   An additional 
shift of +3 km s$^{-1}$ is required to shift the data from the $V_{helio}$
to that of the Local Standard of Rest, $V_{LSR}$.\footnote{We adopt the standard 
{\it IAU} definition of the LSR, assuming a Solar motion of $+20$ km s$^{-1}$
in the direction ($\alpha$, $\delta$)$_{1900}$ = (18$^h$, +30$^{\circ}$) 
[($l$, $b$) $\approx$ (56$^{\circ}$, +23$^{\circ}$)], which gives $V_{LSR}$ = 
$V_{helio}$ + 3 km s$^{-1}$ in the direction of HD~14434.}  The final spectrum 
has a signal-to-noise (S/N) ratio of $\sim$ 40 per resolution element at
1240~\AA.  Further information on the design and performance of STIS
can be found in Woodgate et al. (1998) and Kimble et al. (1998).

\subsection{$FUSE$}
Two exposures of HD~14434 were obtained by $FUSE$ on 1999 November 24 for a 
total exposure time of 4441 seconds (archival IDs P1020504001-002).  The data,
which cover the wavelength range 905 $-$ 1185 \AA, were acquired with the star 
in the center of the large (30~$\!\!^{\prime\prime}$ x 30~$\!\!^{\prime\prime}$)
aperture and resulted in high resolution ($\Delta$$v$ $\sim$ 20 km s$^{-1}$) 
spectra.  The alignment of the four channels was well maintained during both 
exposures.  Only data from the LiF1 channel are presented here since this 
channel has the highest sensitivity.  (This is the channel used for guiding.)  
We utilized the other three channels to provide a consistency check to rule out
the possibility of detector artifacts.  We refer the 
reader to Moos et al. (2000) and Sahnow et al. (2000) for more detailed 
discussions of $FUSE$ and its on-orbit performance.  

The time-tagged data were reduced and calibrated with CalFUSE\footnote{The
CalFUSE pipeline reference guide is available at 
http://fuse.pha.jhu.edu/analysis/pipeline\_reference.html.} (version 1.8.7 
Dixon et al. 2001) the standard $FUSE$ pipeline processing software.
The data were co-added with a cross-correlation technique in order to minimize
uncertainties in the relative wavelength calibration in each spectrum.    The 
wavelength solution used provides good relative calibration across the LiF 
channels.  Relative wavelength calibration errors for $FUSE$ data calibrated 
with CalFUSE v1.8.7 are equivalent to $\sim$ $\pm$ 6-10 km s$^{-1}$ 
(1-$\sigma$).  The absolute wavelength scale of $FUSE$ is not well
constrained.  We used interstellar Fe~{\small II} lines in the STIS and
$FUSE$ spectra to fix the $FUSE$ velocity scale to the STIS $V_{LSR}$ scale.  
The final summed LiF1A spectrum (near 
1050 \AA) and the LiF1B spectrum (near 1140 \AA) have S/N ratios of $\sim$ 20
and $\sim$ 32 per resolution element, respectively.  The uncertainties used in 
calculating the S/N ratios include a significant contribution from fixed-pattern
noise.

\section{Analysis and Results}

\begin{figure}\figurenum{2}\epsscale{0.8}
\begin{center}
\plotone{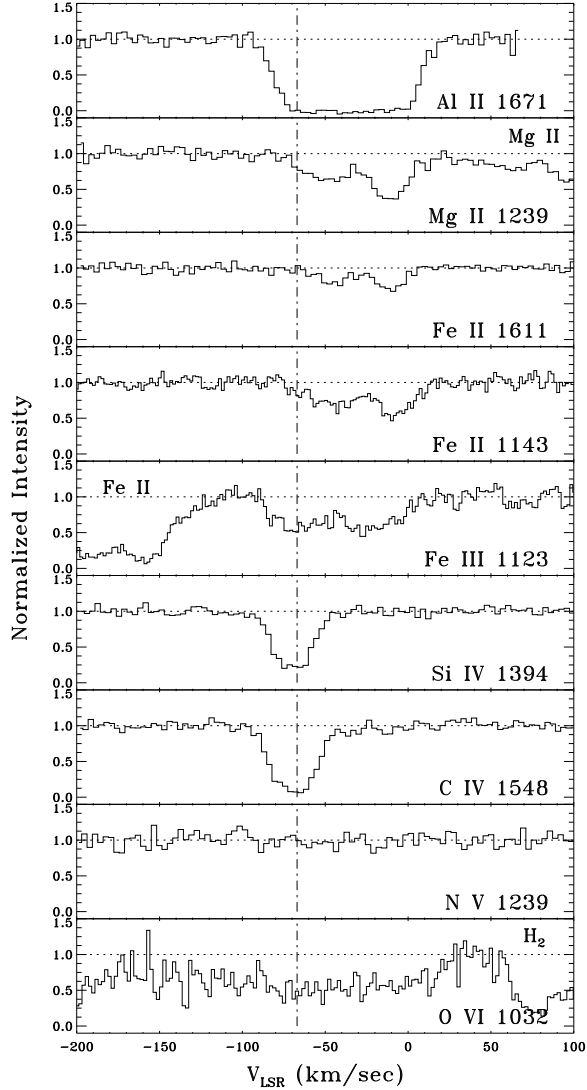}
\vspace{0.5in} 
\caption{\label{IONS}Low and high-ionization absorption along the line of sight
  toward HD~14434 are presented. The \ion{Fe}{2} $\lambda$1143, \ion{Fe}{3}
  $\lambda$1123, and \ion{O}{6} $\lambda$1032 are $FUSE$ data while the 
  remaining spectra were obtained with STIS.  The flux is normalized to unity 
  and is represented by the dotted line.  The dashed-dotted line at -67 km 
  s$^{-1}$ represents the central velocity of the high-velocity highly-ionized 
  gas.  The spectrum containing Al~{\small II} $\lambda$1671 has a cut-off at 
  70 km s$^{-1}$ because it is at the edge of an echelle order on the STIS 
  detector.  Interstellar \ion{C}{2}, \ion{C}{3}, \ion{Si}{2}, and \ion{Si}{3}
  show similar absorption profiles to \ion{Al}{2}.  The broad absorption in the
  \ion{O}{6} spectrum is attributed to a stellar wind feature (see text for 
  details).}
\end{center}
\end{figure}

In our measurements all stellar continua were modeled with low order Legendre 
polynomials (order $\leq$ 4), with the exception of the \ion{O}{6} lines 
(described below). Figure~\ref{IONS} shows normalized absorption profiles of
several low-ionization species (Al~{\small II}, Mg~{\small II}, Fe~{\small II},
and \ion{Fe}{3}) as well as profiles of the high-ionization 
species (\ion{Si}{4}, \ion{C}{4}, \ion{N}{5}, and \ion{O}{6}), which
are the main focus of this work.  As seen from Figure~\ref{IONS}, no detection 
of either \ion{N}{5} or \ion{O}{6} is evident, while \ion{C}{4} and \ion{Si}{4}
are detected only at the velocity of the IVC.  The lower ionization species 
(e.g., \ion{Fe}{2} and \ion{Mg}{2}) are detected primarily in the Local and 
Perseus Arm material.  The Perseus Arm component of \ion{Al}{3} from {\it IUE} 
\citep{SMS01} and \ion{Fe}{3} is blended with the IVC gas.  It is interesting 
to note that no \ion{C}{4} or \ion{Si}{4} absorption is detected at the velocity
of the Local or Perseus Arms suggesting that different physical conditions 
exist for the IVC gas.  We hypothesize that the blue-shifted high-ionization 
IVC is material once associated with the Perseus Arm which has been 
accelerated.    

We use the procedures outlined by Sembach \& Savage (1992) for measuring 
equivalent widths of the Perseus Arm absorption and estimating the errors
in the measurements.  These are given in Table~\ref{ISM_par}, along with 
an equivalent width weighted average velocity.  With the exception of 
\ion{Al}{3} \citep{SMS01}, all measurements are from this work.  The velocities
and column densities measurements agree well from the independent fits from two 
or more lines for each species, except \ion{Fe}{3}.  The difference in 
velocities for the two \ion{Al}{3} lines is due to the poor quality of the 
{\it IUE} data \citep{SMS01}.  The quoted 1-$\sigma$ uncertainties include 
contributions from statistical uncertainties, fluctuations due to fixed pattern
noise, and definable continuum placement errors that were added in quadrature. 
No systematic continuum placement uncertainties were included in the error 
budget.  Uncertainties for the STIS data are dominated by statistical and 
continuum placement uncertainties as the contribution from fixed-pattern noise 
is small.  When no detection was evident, 3-$\sigma$ upper limits are presented 
following the prescription of Morton, York, \& Jenkins (1986).  

\subsection{O~{\small  VI} Concerns}

\begin{figure}\figurenum{3}\epsscale{0.8}
\begin{center}
\plotone{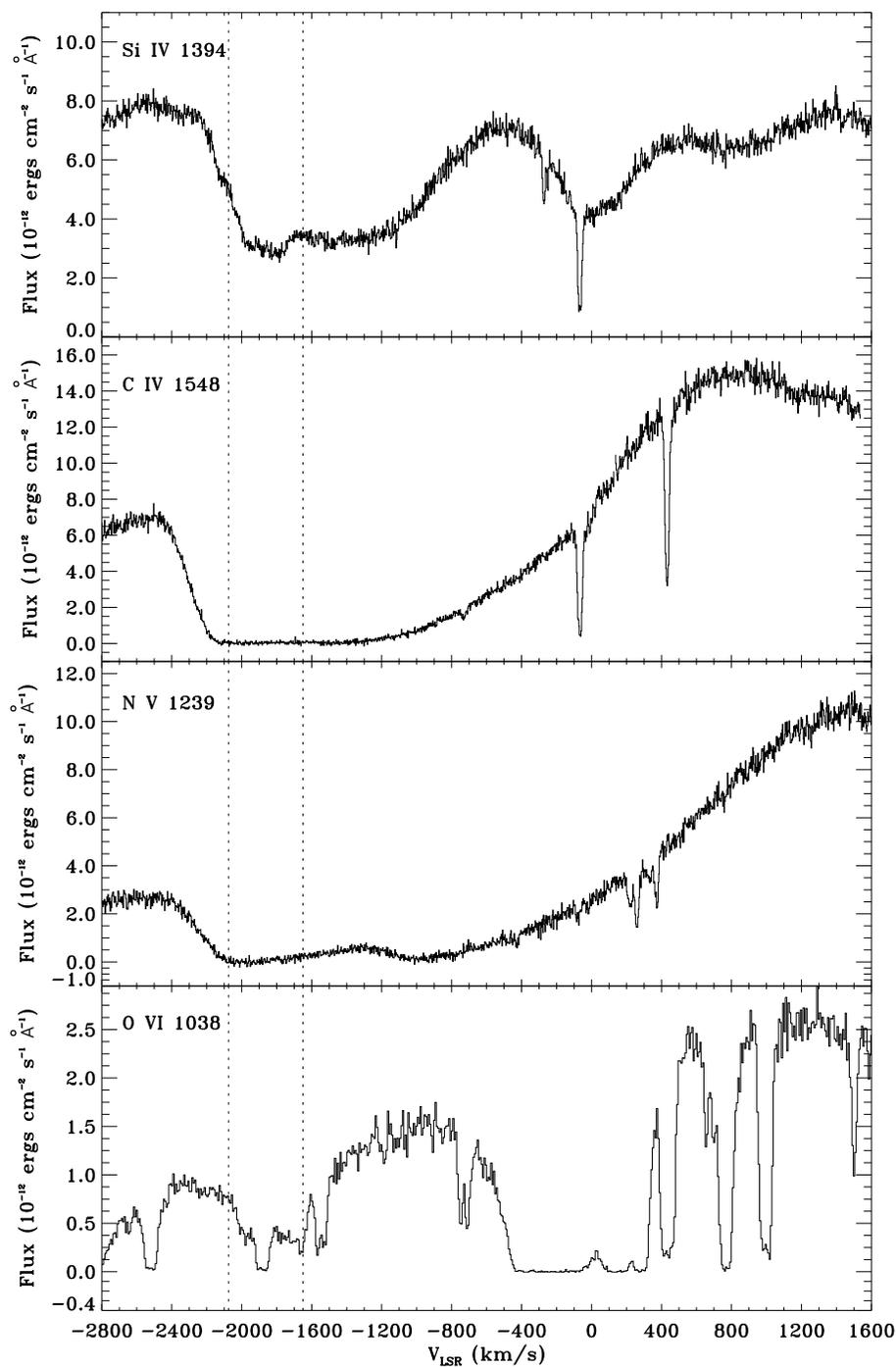}
\vspace{0.5in} 
\caption{\label{winds}
  The wind line profiles of the following species are shown:
  \ion{Si}{4}, \ion{C}{4}, \ion{N}{5}, and \ion{O}{6}.
  There is an obvious wind feature evident in the \ion{Si}{4} wind
  profile.  The dotted lines show the velocity range over
  which the feature is present ($\Delta$$v$ $\sim$ 400 km s$^{-1}$).
  The broad absorption feature in the \ion{O}{6} spectrum
  (bottom panel) is attributed to the stellar wind feature (see text
  for details).  }
\end{center}
\end{figure}

The region around \ion{O}{6} $\lambda$1031.926, shown in the bottom
panel of Figure~\ref{winds}, is complicated by the presence of
\ion{O}{6} in the outflowing stellar wind, H~{\small I}
Ly-${\beta}$ absorption at 1025.723~\AA, and H$_2$ 6$-$0 P(3) and R(4)
absorption at 1031.191 \AA\ and 1032.356 \AA.  The weaker member of
the \ion{O}{6} doublet at 1037.617 \AA\ is not used in the analysis
due to the severe blending with C~{\small II}$^{*}$ at 1037.018 \AA\ 
and the strong H$_2$ 5-0 R(1) and P(1) absorption at 1037.146 \AA\ and
1038.156 \AA.  The stronger member of the \ion{O}{6} doublet may
also be contaminated by HD 6$-$0 R(0) absorption at 1031.912 \AA\ 
(Sembach 1999) along sight lines with substantial H$_2$ [log
$N$(H$_2$) $\geq$ 19].  Careful examination of the entire LiF1A
spectrum reveals the presence of several unblended HD lines arising
from gas in the Local Arm (though none are detected in the Perseus Arm
gas).  We applied a curve of growth analysis (see below) for five
unblended HD lines and determined the equivalent width ($W_{\lambda}$)
of the contaminating HD line to be 58.0 m\AA.  A simple Gaussian model
of the HD contamination was generated and divided into the normalized
\ion{O}{6} profile to produce the final unblended \ion{O}{6}
spectrum.

The continuum placement of the stronger interstellar \ion{O}{6}
line is potentially affected by the presence of a stellar wind feature
in the weaker \ion{O}{6} line.  This feature is not detected in the
stronger \ion{O}{6} line due to its overlap with H~{\small I}
Ly-$\beta$.  Figure~\ref{winds} shows the stellar wind profiles of
\ion{Si}{4} $\lambda$1393.755, \ion{C}{4} $\lambda$1548.195,
\ion{N}{5} $\lambda$1238.821, and \ion{O}{6} $\lambda$1031.926. A
stellar wind feature is clearly observed in \ion{Si}{4} over the range
-2100 km s$^{-1}$ $\leq$ $v$ $\leq$ -1700 km s$^{-1}$ (see top panel
of Figure~\ref{winds}), close to the terminal velocity of V$_{\infty}$
= 2120 $\pm$ 20 km s$^{-1}$ (this work), which is $\sim$ 200 km s$^{-1}$ greater
than that derived by Howarth \& Prinja (1989) based on {\it IUE} data.  The 
terminal wind velocity is determined from the short wavelength absorption limit
of the P~Cygni \ion{C}{4} wind line at 1548 \AA.  This wind feature cannot
be seen in either the \ion{C}{4} or \ion{N}{5} winds because they
are saturated (see middle two panels of Figure~\ref{winds}).  The wind
feature coincides in velocity space with a broad depression near the
interstellar \ion{O}{6} line (demarcated by the vertical dashed lines in
Figure~\ref{winds}).  Though the STIS and $FUSE$ data were acquired about
two years apart, stellar wind features that occur at or near the
terminal velocity, as is the case here, are typically long-lasting
entities (A. Fullerton 2002, private communication).  These factors
indicate that the broad absorption coincident with the interstellar
\ion{O}{6} 1032 \AA\ absorption could be due to structure in the
outflowing wind of the star.  Due to the uncertainties in modelling the
stellar continuum in the region surrounding the \ion{O}{6} 1032
\AA\ absorption, we give a conservative upper limit to $N$(\ion{O}{6}) in 
Table~\ref{ISM_par}.

\subsection{Apparent Optical Depth Method} 

A useful tool for determining the column density of an absorbing
species is the apparent optical depth method (Savage \& Sembach 1991).
The apparent optical depth, $\tau_a(v)$, is a valid,
instrumentally-blurred representation of the the true optical depth in
the absence of unresolved saturated structure within the line
profile.  The continuum normalized absorption line profile, $I(v)$
$\equiv$ $e^{-\tau_a(v)}$, is related to apparent column density per
unit velocity, $N_a(v)$, by

\begin{equation}
N_a(v) = {{m_ec} \over {\pi e^2}}~{{\tau_a(v)} \over {f\lambda}} = 3.768 \times
10^{14}~{{\tau_a(v)} \over {f\lambda ({\rm \AA})}}~{\rm ions~cm}^{-2}~({\rm 
km~s}^{-1})^{-1}.
\end{equation}  

\noindent In this equation, $m_e$ and $e$ denote the mass and charge of the 
electron, respectively, $c$ is the speed of light, $f$ is the atomic
oscillator strength, and $\lambda$ is the laboratory wavelength of the
transition.  Savage \& Sembach (1991) showed that $N_a(v)$ is
equivalent to the true column density as a function of velocity,
$N(v)$, if there is no unresolved saturated structure.  The presence
of unresolved saturated structure can be identified through the
comparison of $N_a(v)$ profiles for different transitions of the same
species with differing $f$-values.  In situations where unresolved saturation
is present, the derived $N_a(v)$ is always a lower limit to the true column 
density.  The integrated apparent column densities are listed in 
Table~\ref{ISM_par}.

\begin{figure}\figurenum{4}\epsscale{0.8}
\plotone{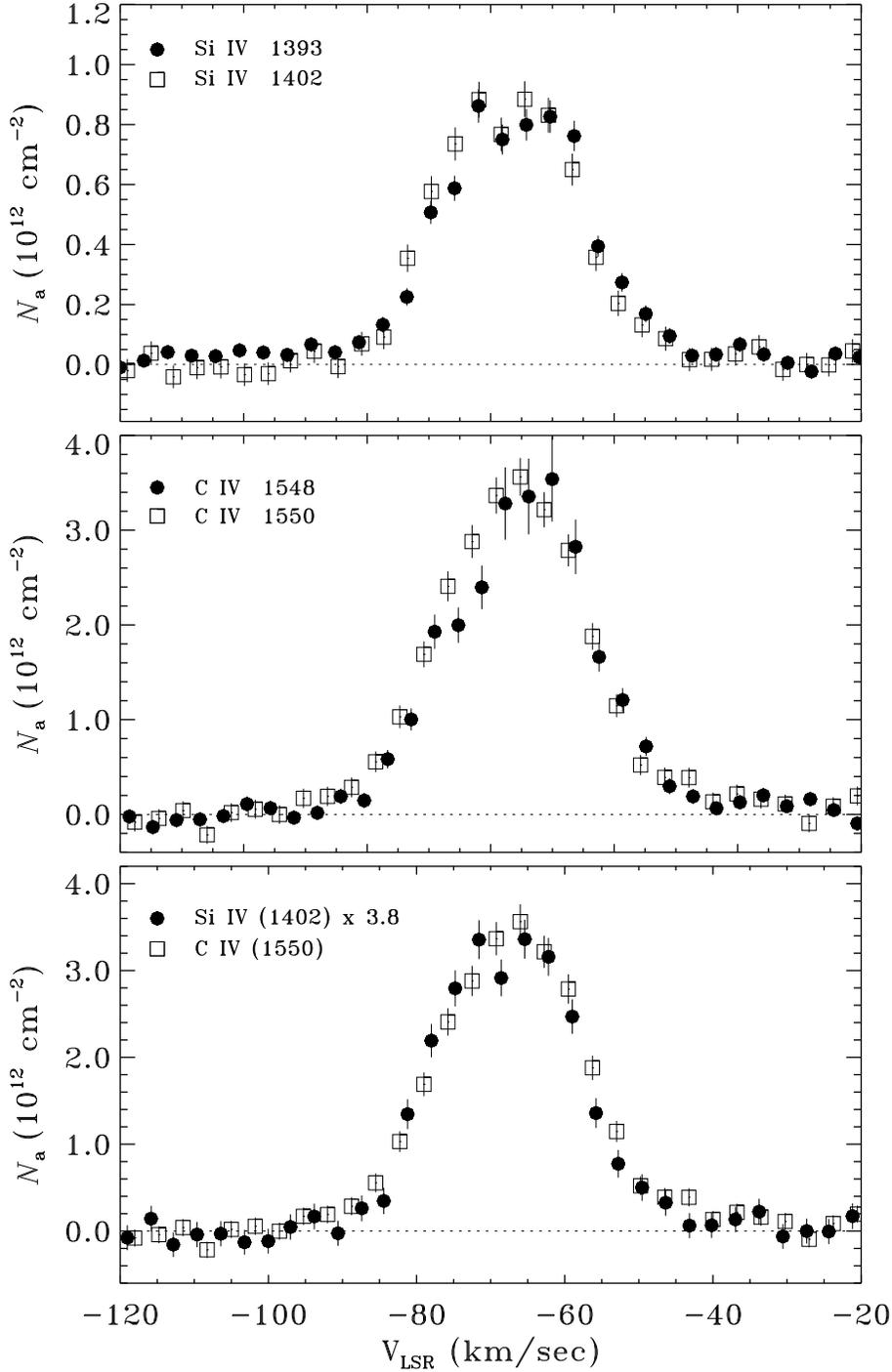}
\vspace{0.3in} 
\caption{\label{AOD} {\it Upper panel:} $-$ Apparent
column density profiles of the \ion{Si}{4} 1393 and 1402~\AA\ lines.
The solid circles represent the stronger member of the doublet, while the
open squares denotes the weaker line.  {\it Middle panel:} $-$ Apparent column 
densities of the \ion{C}{4} doublet $\lambda\lambda$ 1548, 
1550.  The symbols are the same as for the top panel.  {\it Bottom panel:}$-$
Apparent column denisties of the weaker members of the \ion{Si}{4} and \ion{C}{4} doublets and
reveals that a \ion{C}{4}/\ion{Si}{4} ratio of 3.8 is representative over 
the entire velocity range.  The overlap of both species suggests that they
are co-spatial.  There is excellent agreement in the $N_a$($v$) of both members 
of the \ion{Si}{4} and \ion{C}{4} doublets indicating little or no 
evidence of saturation.}
\end{figure}

The $N_a(v)$ profiles for both members of the \ion{Si}{4} and
\ion{C}{4} doublets are shown in the top and middle panels,
respectively, of Figure~\ref{AOD}.  In each case the stronger line is
denoted by the solid circles, while the weaker line is represented by
the open squares.  The overall agreement between the $N_a(v)$ profiles
for both members of the \ion{Si}{4} and \ion{C}{4} doublets
implies little or no evidence for saturation, and hence $N_a(v)$ =
$N(v)$.  The bottom panel of Figure~\ref{AOD} shows a comparison of
the weaker members of the \ion{Si}{4} and \ion{C}{4} doublets
along the line of sight toward HD~14434.  There is an excellent
correspondence between the \ion{Si}{4} and \ion{C}{4} profiles
over the entire velocity extent of the lines.  This similarity
suggests that \ion{Si}{4} and \ion{C}{4} arise in the same gas
(i.e., they are co-spatial).  In addition, the similar line widths
indicate that the dominant broadening mechanism is non-thermal (e.g.,
turbulence).  The derived \ion{C}{4}/\ion{Si}{4} ratio is 3.8
$\pm$ 0.3, quite similar to the Galactic average of 4.3 $\pm$ 1.9
(Sembach, Savage, \& Tripp 1997).

As a check on the apparent optical depth method, we performed curve of growth
(COG) analyses for HD and \ion{Fe}{2}.  All HD lines are on the 
flat part of the curve of growth and no HD column density was obtained.
The COG for \ion{Fe}{2} yielded log[$N$(\ion{Fe}{2})] = 14.90 $\pm$ 0.06 
cm$^{-2}$ and a $b$-value of 23.0 $\pm$ 4.4 km s$^{-1}$; the large $b$-value is
due to the contribution of both IVC and Perseus Arm material.  This column 
density is in good agreement with that determined from the apparent optical 
depth method, therefore \ion{Fe}{2} is unsaturated.  The unsaturated \ion{Fe}{2}
lines (see Table~\ref{ISM_par}) probe a larger apparent optical depth 
($\tau_a$ $\leq$ 3) than does \ion{Fe}{3} $\lambda$1122.546 ($\tau_a$ $\leq$ 1).
Subsequently, we assume that \ion{Fe}{3} is unsaturated.  We adopt the curve of 
growth results for the \ion{Fe}{2} and the apparent column density results for 
the \ion{Mg}{2}, \ion{Fe}{3}, \ion{Si}{4}, and \ion{C}{4}. The final adopted 
columns are presented in Table~\ref{ISM_den}.  No results are tabulated for HD 
due to the fact that all HD lines are on the flat part of the curve of growth.

\subsection{Component Synthesis}
\begin{figure}\figurenum{5}\epsscale{0.8}
\plotone{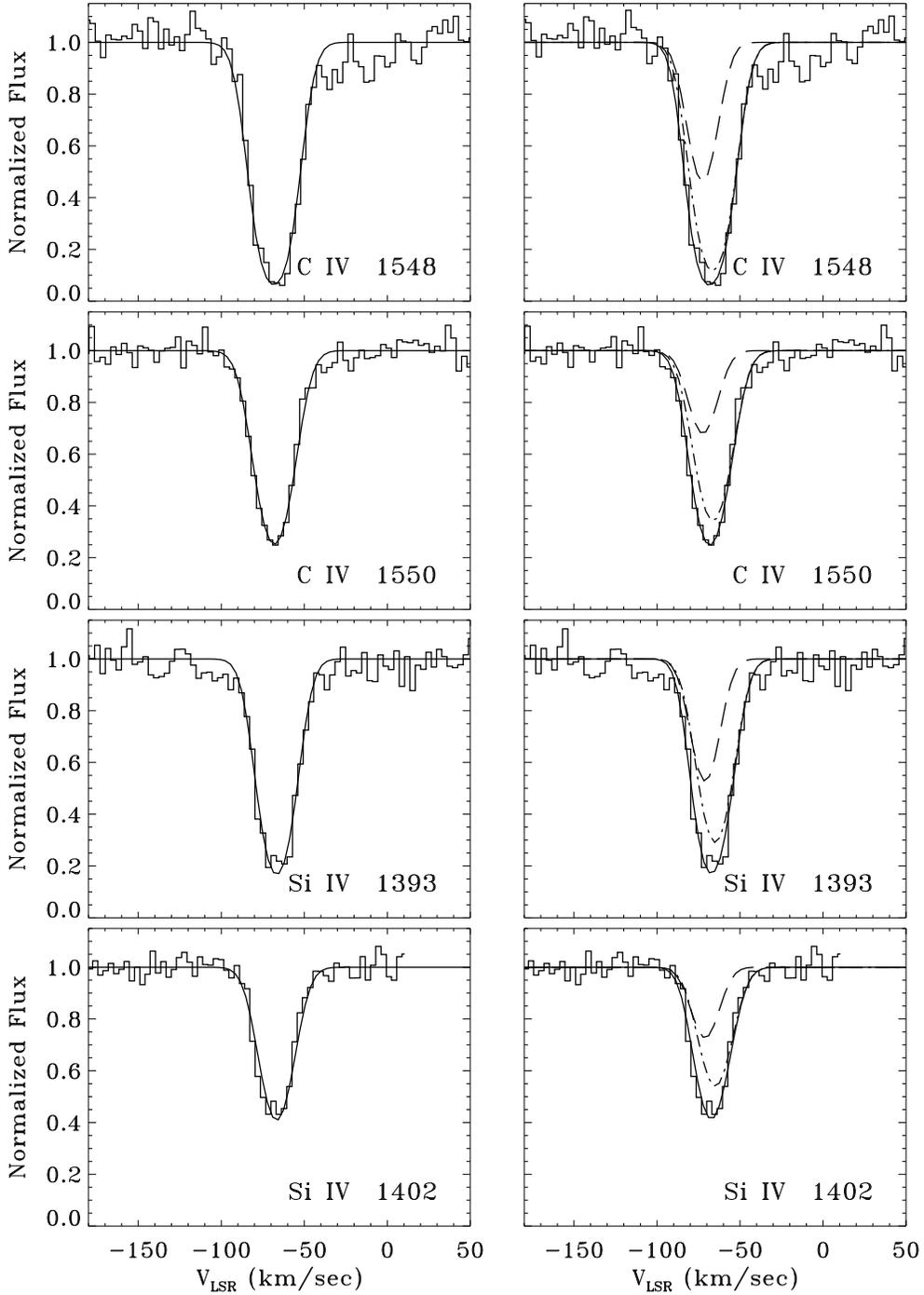}
\vspace{0.5in} 
\caption{\label{1+2comps}  The left panels show the one-component model
fits to the \ion{Si}{4} and \ion{C}{4} doublets.  The right side shows 
the two-component model fits.  There is only a 30\% confidence in the 
second component.}
\end{figure}

A visual examination of the \ion{Si}{4} and \ion{C}{4} line
profiles in Figure~\ref{IONS} reveals a slight asymmetry in the line
profiles suggesting the presence of multiple velocity components.  We,
therefore, used the profile fitting code {\bf Owens.f} developed by
Martin Lemoine and the French {\it FUSE} Team (Lemoine et al. 2002) to
fit both one- and two-component model to the data.  Because the
\ion{Si}{4} and \ion{C}{4} absorption toward HD~14434 are
co-spatial, the line profiles of both species were fit simultaneously
keeping only the number of velocity components fixed.  The free
parameters in the profile synthesis are: $V_{LSR}$, $N$(\ion{C}{4}),
$N$(\ion{Si}{4}), nonthermal velocity ($v_{nt}$), and
kinetic temperature ($T$).  The values of $v_{nt}$ and $T$ were
obtained from the following definition of the Doppler broadening
parameter, $b^2$ = 2$k$$T$/$Am_H$ + $v_{nt}^2$, where $k$ is Boltzmann's 
constant, $A$ is the atomic weight, and $m_H$ is the mass of a 
hydrogen atom.  The fits are shown
in Figure~\ref{1+2comps} and the results summarized in
Table~\ref{CIV_SiIVcomponents}.  The quoted 1-$\sigma$ uncertainties
do not include systematic errors in the continuum placement, although
definable statistical and continuum placement errors are included (see Sembach
\& Savage 1992).

The best one-component model resulted in a reduced $\chi^2$
($\chi^2$/$\nu$) of 1.42.  For this model, a value of $v_{nt}$ =
12.0~$\pm$~0.2 km s$^{-1}$ and $T$ = 10,450~$\pm$~3,400 K are derived.
For the two-component model, the values of $v_{nt}$ and $T$ are 10.7
km s$^{-1}$ and 26,500 K and 11.3 km s$^{-1}$ and 10,100 K, for the blue
and red velocity components, respectively.  The two-component fit
resulted in an increased $\chi^2$/$\nu$, compared with the
one-component model.  The application of an F-test (Lupton
1993) to the model results reveals that the second component is
justified at the 30\% confidence level.  Thus, the evidence for a
second component is not statistically significant.  Due to its low
confidence, no error bars are given for the two component model. 

\section{The Nature of the Highly-Ionized, Intermediate-Velocity Cloud}

The strong \ion{Si}{4} and \ion{C}{4} absorption toward HD~14434
is found in a single IVC centered at -67
km s$^{-1}$ that is offset by $\sim$ -20 km s$^{-1}$ from the
lower-ionization absorption associated with the Perseus Arm gas (see
Figure~\ref{IONS}).  A profile synthesis analysis shows this
highly-ionized IVC is dominated by
nonthermal broadening with $v_{nt}$ = 12.0 $\pm$ 0.2 km s$^{-1}$ and
is at a relatively cool temperature of $T$ = 10,450 $\pm$ 3,400 K.  Our
apparent optical depth analysis (see Figure~\ref{AOD}) suggests that
\ion{Si}{4} and \ion{C}{4} are co-spatial along the line of sight
with a \ion{C}{4}/\ion{Si}{4} ratio of 3.8 $\pm$ 0.3, which is consistent with
the Galactic average (Savage, Sembach, \& Tripp 1997).  Our measured ratio is 
similar to \ion{C}{4}/\ion{Si}{4} $\sim$ 3 found by McLaclan \& Nandy (1985)
using {\it IUE} data.  They concluded that a supernova explosion was the origin 
of the intermediate-velocity \ion{C}{4} and \ion{Si}{4} absorption
toward Per~OB1.  However, our derived low temperature does not support
this conclusion.  Therefore, in this section we discuss the existing
models as they apply to the origin of the high-ion
intermediate-velocity absorption toward HD~14434.

At a temperature of 10$^4$~K, neither \ion{Si}{4} or \ion{C}{4}
would be present if equilibrium collisional ionization prevailed
(Sutherland \& Dopita 1993).  It is generally believed that the
presence of \ion{Si}{4} and \ion{C}{4} absorption in the ISM
often arises as a result of non-equilibrium collisional ionization at
interfaces of hot gas ($T$ $\sim$ 10$^6$~K) with relatively cool ($T$
$\sim$ 10$^2$ $-$ 10$^4$ K) interstellar clouds.  Spitzer (1996)
provides a good, concise review of the existing models.
Non-equilibrium collisional ionization, through models of turbulent
mixing layers (Slavin, Shull, \& Begelman 1993), cooling galactic
fountain gas (Shapiro \& Benjamin 1992), and
conductive interfaces (Ballet, Arnaud, \& Rothenflug 1986;
B\"{o}hringer \& Hartquist 1987; Borkowski, Balbus, \& Fristrom 1990;
Breitschwerdt \& Schmutzler 1994), results in gas at $T$ $\sim$
10$^5$~K and cannot reproduce the observed \ion{C}{4}/\ion{Si}{4}
ratio.  Inclusion of self-photoionization (Shapiro \& Benjamin 1992;
Slavin et al. 1993; Breitschwerdt \& Schmutzler 1994) improves the
agreement between the observed \ion{C}{4}/\ion{Si}{4} ratio and
that predicted by the models.  However, these models simultaneously
over-produce both \ion{N}{5} and \ion{O}{6} by an order of
magnitude and have an order of magnitude higher temperature compared
with the highly-ionized IVC toward HD~14434.  In general, models which
rely solely on collisional ionization, including turbulent mixing
layers, cooling galactic fountain gas, and conductive interfaces,
cannot explain the \ion{Si}{4} or \ion{C}{4} absorption toward
HD~14434.

In a time-dependent non-equilibrium model of radiative cooling, 
Shapiro \& Moore (1976) calculated the ionization fraction of several 
species (H, O, N, C, Si, etc.).  In non-equilibrium situations, gas 
initially at $T \geq 10^5$ K would cool more rapidly than high-ions recombine
leaving the gas ``over ionized'' for its temperature (Kafatos 1973). Their 
model follows a parcel of cooling hot gas from an initial temperature of 
10$^6$ K to a final temperature of 10$^4$ K.  The lower temperature is similar 
to that derived for the highly-ionized IVC toward HD~14434.  The resulting 
\ion{C}{4}/\ion{Si}{4} and \ion{C}{4}/\ion{Fe}{3} ratios predicted by Shapiro 
\& Moore (1976) at $T$ $\sim$ 10$^4$ K are 1.6 and 7.4.  The
\ion{C}{4}/\ion{Si}{4} ratio is approximately a factor of 2 too low, while the 
\ion{C}{4}/\ion{Fe}{3} ratio is consistent with the observations.  More
recently, Edgar \& Chevalier (1986) developed a model similar to that
of Shapiro \& Moore (1976) which included updated atomic data for more
species and followed the cooling gas with several assumptions regarding its 
evolution (e.g., isobaric, isochoric).  Their calculations resulted
in a \ion{C}{4}/\ion{Si}{4} ratio is a factor of 3 to 10 times
larger (depending on the assumptions) than the observed ratio toward
HD~14434.  Inclusion of self-photoionization may resolve the differences 
between these models and the observations.  Future non-equilibrium calculations 
which include the effects of self-photoionization are needed to determine if 
this is a viable mechanism to explain the \ion{C}{4}/\ion{Si}{4} ratio.  

Another model which results in the production of \ion{Si}{4} and
\ion{C}{4} at relatively cool temperatures is that of wind-blown
interstellar bubbles, whose evolution was outlined by Castor, McCray, \& 
Weaver (1975) and Weaver et al. (1977).  Stars of spectral type earlier than
B2 are known to have strong stellar winds (Howarth \& Prinja 1989).  The 
outflowing stellar wind of a star impinges on the ambient ISM and can create
an interstellar bubble which have been found to be common around
Wolf-Rayet and O-type stars (Chu et al. 1982; Lozinskaya 1991; Cappa \& 
Herbstmeier 2000).   The Weaver et al. (1977) model predicts the
physical quantities associated with shells driven by the winds from
early-type stars including the radius of the shell, expansion velocity, and the
column density of \ion{O}{6} and other highly-ionized species.  
Comparison of the
high-ion predictions of the Weaver et al.  model\footnote{We adopt the stellar
parameters listed in Table~\ref{star_par} and the following parameters for the 
model: an ambient interstellar density, $n_o$ = 1 cm$^{-3}$, and a
typical age of an interstellar bubble, $t = 2.3$ Myr (Cappa \&
Herbstmeier 2000).} with the properties of the highly-ionized IVC
toward HD~14434, assuming it is associated with a wind-blown bubble,
results in a \ion{C}{4}/\ion{Si}{4} ratio $\sim$ 16 which is a
factor of 4 too large.  Therefore, we conclude that the presence of a
Weaver-type bubble cannot be responsible for the observed
highly-ionized IVC toward HD~14434.  

The derived temperature ($T$ $\sim$ 10$^4$~K) and turbulent
velocity ($v_{nt}$ $\sim$ 12 km s$^{-1}$) of the IVC are
typical of gas found in Galactic H~{\small II} regions (Reynolds
1991).  The low temperature, in particular, suggests that
photoionization may play a role in the origin of the high ions toward
HD~14434.  We used the photoionization equilibrium code Cloudy (v94;
Ferland 1996; Ferland et al. 1998) to invesigate the possibility that
photoionization may explain the observed properties of the IVC seen
toward HD~14434.  We considered two photoionization scenarios: ($i$) the
IVC is associated with a normal \ion{H}{2} region surrounding HD~14434;
($ii$) the IVC has been ionized by emission from a hot, thermal
plasma.

Following Howk \& Savage (1999), a suite of single-star \ion{H}{2} region models
were computed using Cloudy and the stellar parameters in Table~\ref{star_par}. 
For simplicity, we assume that X-rays emitted by the stellar wind can be 
modelled as an equilibrium plasma at $T = 10^6$~K with an 
intensity normalized such that $L_X/L_{bol} \sim 10^{-7}$~K (e.g., Cassinelli 
et al. 1981).  The results of these models are summarized in Table \ref{CLOUDY}.
All the models yield a predicted \ion{C}{4}/\ion{Si}{4} ratio $\leq1$.  This
ratio could be raised by assuming lower densities, but the lowest density model
in Table~\ref{CLOUDY} should already be considered extreme.  We note that these
models, following Sembach et al. (2000), assume depleted abundances appropriate 
for the warm neutral medium (WNM) of Galaxy; assuming solar system abundances would 
lower this ratio.  Thus standard \ion{H}{2} region models seem unable to explain
the observed properties of the gas.

\begin{figure}\figurenum{6}
\rotatebox{90}{\epsscale{0.8}
\plotone{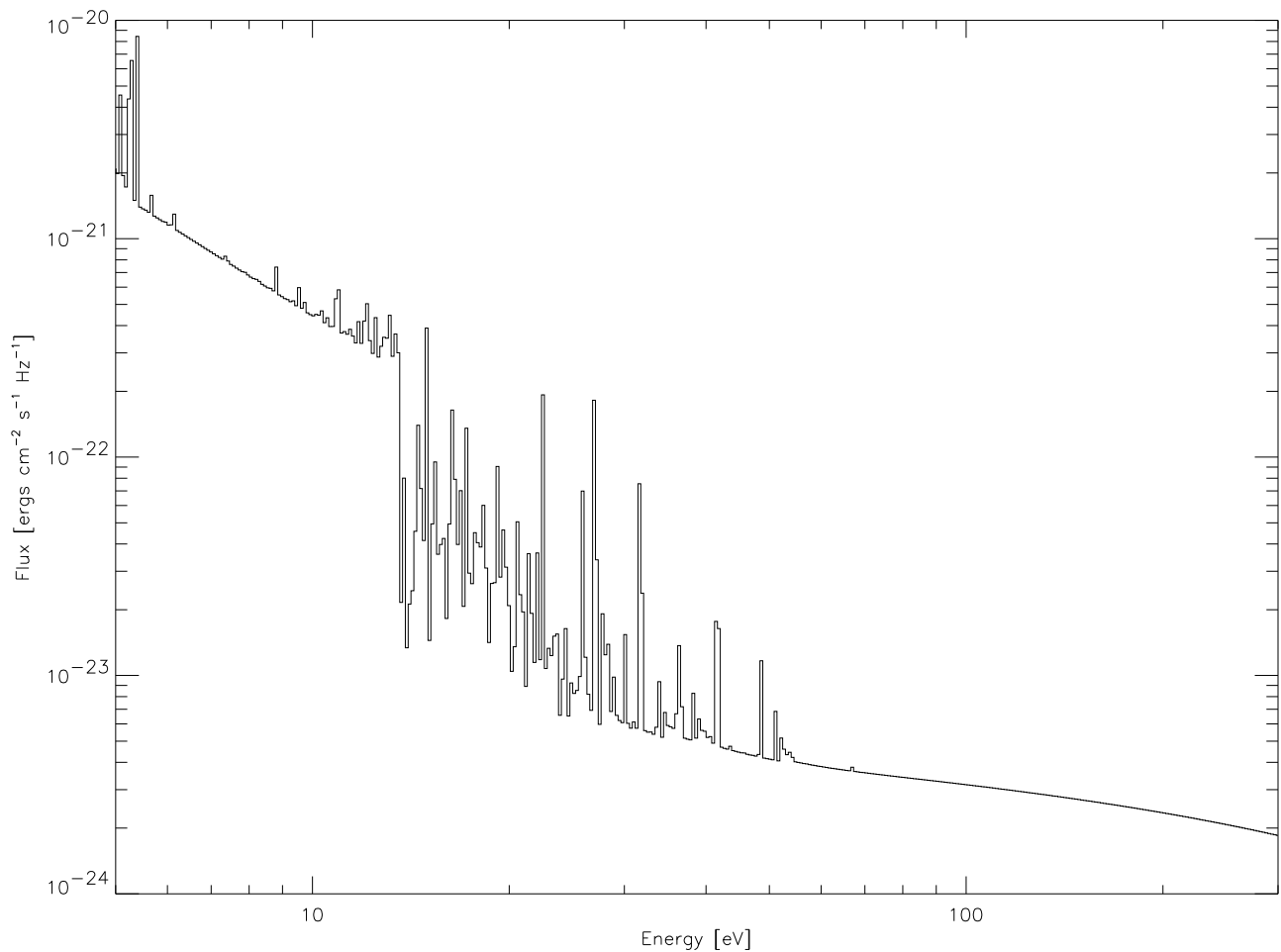}}
\vspace{0.5in} 
\caption{\label{ion_input}  The ionizing spectrum from cooling hot diffuse gas
that was used to calculate Models 1 and 2 in Table 6 is shown.  This spectrum 
includes the diffuse interstellar radiation field (Black 1987) and emission 
from a thermal plasma at $\log(T_{\rm plasma}) = 6.1$~K whose flux was normalized to the observed 
diffuse X-ray background for the Local Bubble.}
\end{figure}

Finally, we investigated the ionization characteristics of a plane-parallel 
cloud illuminated by emission from a diffuse hot plasma.  We assume an 
ionizing spectrum made up of a diffuse interstellar radiation field 
(Black 1987) and, following Slavin \& Frisch (2002), emission from a thermal
plasma.  The thermal plasma is chosen to have a temperature of
$\log(T_{\rm plasma}) = 6.1$~K (Models 1, 2, and 3) or 7.0~K (Model 4).  For 
Models 1, 2, and 4, the flux is normalized to match the observation of the 
diffuse X-ray background of the Local Bubble.  For Model 3, we normalized the
emission spectrum to a diffuse X-ray background with a factor-of-10 higher flux.
Figure~\ref{ion_input} shows the spectrum of ionizing radiation was used as the
source of photoionization for our Cloudy Models 1 and 2.  We have investigated 
the use of the Solar System abundances (Model 1; Anders \& Grevesse 1989) as 
well as the depleted abundances characteristics of the WNM\footnote{The Solar 
abundance model results in a lower temperature because interstellar grains were 
not included.  A more detailed treatment of interstellar grains results in 
approximately a factor-of-2 increase in the temperature.} described above
(Models 2, 3, and 4; following Sembach et al. 2000).  We then calculated the 
combination of total hydrogen density and column density that provided the 
closest match to the \ion{C}{4} and \ion{Si}{4} column densities seen in the IVC
toward HD~14434.

Table \ref{localbubble} summarizes the properties of our photoionization models
of a gas cloud illuminated by emission from a cooling hot plasma. 
All models match the observed 
\ion{C}{4}/\ion{Si}{4} ratio for reasonable $N$(H) of about 10$^{19}$ cm$^{-2}$.
The resulting temperatures are within 2-$\sigma$ of that derived from the 
\ion{Si}{4} and \ion{C}{4} line widths.  Models 1 and 2 predict densities and 
temperatures of $n_{\rm H}$ = 0.001 cm$^{-3}$ and $T$ = 4,630 K and $n_{\rm H}$
0.002 cm$^{-3}$ and $T$ = 6,930 K, respectively.  Increasing the X-ray flux 
(Model 3) results in $n_{\rm H}$ = 0.016 cm$^{-3}$ and $T$ = 10,100 K and 
increasing the temperature of the emitting plasma (Model 4) results in
$n_{\rm H}$ = 0.2 cm$^{-3}$ and $T$ = 6,880 K.  The \ion{Si}{4} and \ion{C}{4}
column densities predicted by all the models in Table~\ref{localbubble} agree 
with the observations to better than 5\%.  In addition, these models do not 
over predict the column densities of \ion{N}{5} or \ion{O}{6}.  With the 
exception of Model 3, the predicted \ion{Al}{3}/\ion{Fe}{3} ratio does not 
match the observations because the IVC absorption is blended with the Perseus 
Arm absorption.  However, the predicted \ion{C}{4}/\ion{Al}{3} and 
\ion{C}{4}/\ion{Fe}{3} ratios are in agreement with the lower limits obtained 
from the observations, therefore our models do not overproduce either 
\ion{Al}{3} or \ion{Fe}{3}.  It is important to note that 
the values quoted for \ion{C}{4}/\ion{Al}{3} and \ion{C}{4}/\ion{Fe}{3}
assume that Fe and Al are depleted heavily in the models using the depleted 
abundances of the WNM.

Models 1, 2, and 4 result in P/k $\leq$ 340 K cm$^{-3}$, significantly lower 
than the average pressure for interstellar clouds P/k $\sim$ 2500 K cm$^{-3}$ 
(Jenkins \& Tripp 2001).  Lower pressures are not unexpected because models of
$nT$ for warm neutral cloud halos (Andersson \& Wannier 1993) predict $nT$ $<$
1000 K cm$^{-3}$ at the height of Per~OB1 above the Galactic plane.  We note 
that the X-ray flux could be stronger in this active region of the Galaxy than
in the solar neighborhood.  Our Model 3 investigates such an increase in the 
X-ray flux and results in P/k = 2900 K cm$^{-3}$, as expected from Jenkins 
\& Tripp (2001).  Model 3 also yields a temperature which is in agreement with 
that observed for the IVC.  Note: increasing the density in Model 2 to 
$\log(n_{{\rm H}})$ = -2.2 cm$^{-3}$ and $N$(H$_{tot}$) = 20.4 cm$^{-2}$ 
without increasing the radiation field results in a significantly lower 
\ion{C}{4}/\ion{Si}{4} = 2.0 for the observed $N$(\ion{C}{4}).

There is excellent agreement between our hot plasma photoionization models 
and the observations of the IVC toward HD~14434.  Thus, emission of ionizing
radiation by cooling hot gas is likely to be the source of the ionization in
this cloud.  The cooling radiation hypothesized as the ionization source in 
these models could arise from a general diffuse hot-ionized medium or from the 
super-heated interior of a supershell in the Perseus Arm (e.g., GS139-03-69).

Given the similarity of the \ion{C}{4}/\ion{Si}{4} ratio in this IVC to the 
Galactic average, an extension of our results implies that much of the Galactic
\ion{C}{4} and \ion{Si}{4} absorption could arise as a result of 
photoionization by cooling radiation from a hot, thermal plasma.  We note that
\citet{F03} have identified narrow (warm) components in \ion{C}{4} and 
\ion{Si}{4} toward the star HD~116852, suggesting that such 
absorption is not uncommon in the Galaxy (although their \ion{C}{4}/\ion{Si}{4}
ratio was significantly lower than that seen toward HD~14434).
Further analyses of high-resolution observations of the highly-ionized ISM are
needed to determine if this phenomenom is common throughout the Galaxy.

\section{Summary}

Our STIS observations of HD~14434 reveal an IVC detected in
\ion{C}{4} and \ion{Si}{4} and possibly in \ion{Fe}{3} and \ion{Al}{3}
\citep{SMS01}.  The intermediate-velocity \ion{Si}{4} and
\ion{C}{4} absorption toward HD~14434 is contained in a single velocity
component at a temperature of 10,450 $\pm$ 3,400 K and with a non-thermal
dispersion of 12.0 $\pm$ 0.2 km s$^{-1}$, similar to the values found for
Galactic H~{\small II} regions.  The observed high-ion column
densities are log[$N$(\ion{Si}{4})] = 13.34 $\pm$ 0.02 cm$^{-2}$ and
log[$N$(\ion{C}{4})] = 13.92 $\pm$ 0.02 cm$^{-2}$, which yield a \ion{C}{4}/\ion{Si}{4}
ratio of 3.8 $\pm$ 0.3 that is remarkably similar to the Galactic average of 
4.3 $\pm$ 1.9 (Sembach, Savage, \& Tripp 1997).

The presence of \ion{Fe}{3}, \ion{Al}{3}, \ion{Si}{4} and \ion{C}{4} in 
low-temperature (10$^4$~K) gas suggests that these species are in a
region dominated by photoionization.  The lack of significant
\ion{N}{5} and \ion{O}{6} absorption also support photoionization
as the source of the highly-ionized IVC.  Our photoionization model
utilizing emission from a diffuse hot plasma is in excellent agreement with our
observations indicating that this is the origin of the IVC toward HD~14434.  In
addition, the similarity of our \ion{C}{4}/\ion{Si}{4} ratio with the Galactic 
average suggests that much of the \ion{Si}{4} and \ion{C}{4} detected in the 
ISM could be produced from ionizing photons from a cooling hot gas.

\acknowledgments We would like to thank the dedicated {\it FUSE}
mission planners for their efforts in planning these observations.
This work is based on data obtained for the Guaranteed Time Team by
the NASA-CNES-CSA $FUSE$ mission operated by the Johns Hopkins
University.  Financial support to U.S.  participants has been provided
by NASA contract NAS5-32985.  The Wisconsin H-Alpha Mapper is funded by 
the National Science Foundation.  This research made use of the Simbad
database, operated at CDS, Strasbourg, France.

\clearpage
\newpage

\clearpage
\newpage

\newpage
\clearpage

\clearpage
\newpage

\clearpage
\newpage

\clearpage

\newpage
\setcounter{table}{0}
\begin{deluxetable}{cccccccccccc}
\tablecolumns{3}
\tablewidth{0pt}
\tablecaption{Adopted Stellar Parameters for HD~14434 \label{star_par}}
\tablehead{ \colhead{Quantity} & \colhead{Value} & \colhead{Source} } 
\startdata
Spectral Type & O5.5 Vnfp & Walborn (1972) \\
$V$ [mag] & 8.49 & Simbad\tablenotemark{a} \\
$E(B-V)$ [mag] & 0.48 & Diplas \& Savage (1994) \\
($l$, $b$) & (135.$\!\!^{\circ}$1, -3.$\!\!^{\circ}8$) & 
Simbad\tablenotemark{a} \\
$T_{eff}$ [K] &  44,000 & Howarth \& Prinja (1989) \\
log($L_{*}/L_{\odot}$) & 5.4 & Howarth \& Prinja (1989) \\
$V_{\infty}$ [km s$^{-1}$] & 2,120 & This work \\
$d$ [kpc] & 2.3 & Diplas \& Savage (1994) \\
$z$ [pc] & -153 & Diplas \& Savage (1994) \\
$N$(H~{\small I}) [cm$^{-2}$] & (2.82 $\pm$ 1.21) $\times$ 10$^{21}$ & 
Diplas \& Savage (1994) \\
\enddata
\tablenotetext{a}{The Simbad database is operated at CDS, Strasbourg, FR.}
\end{deluxetable}
\clearpage

\setcounter{table}{1}
\begin{deluxetable}{cccccccc}
\tablecolumns{7}
\tablewidth{0pt}
\tablecaption{Ions toward HD~14434 \label{ISM_par}}
\tablehead{ 
\colhead{Species} & \colhead{IP$^{~{\rm a}}$} & \colhead{$\lambda_o^{~{\rm b}}$} 
 & \colhead{log{($\lambda$$f$)$^{~{\rm b}}$}} & \colhead{$V_{LSR}$} & 
\colhead{$W_{\lambda}$$^{~{\rm c}}$} & \colhead{log($N_{\rm a}$)$^{~{\rm d}}$} \\
\colhead{} & \colhead{[eV]} & \colhead{[\AA]} & \colhead{[\AA]} 
 & \colhead{[km s$^{-1}$]} & \colhead{[m\AA]} & \colhead{[cm$^{-2}$]} }
\startdata
Mg {\small II} & 7.7$-$15.0 & 1239.925 & 0.190$^{{\rm e}}$ & -47.1 $\pm$ 1.1 & 
43.1 $\pm$ 3.7 & 15.77 $\pm$ 0.04 \\
 & & 1240.395 & -0.110$^{{\rm e}}$ & -48.1 $\pm$ 2.2 & 19.9 $\pm$ 3.8 & 
 15.67 $\pm$ 0.09 \\
 & & & & & &  \\
Fe {\small II} & 7.9$-$16.2 & 1121.975 & 1.351$^{{\rm f}}$ & -46.2 $\pm$ 0.9 & 
140.6 $\pm$ 5.0 & 15.05 $\pm$ 0.03  \\
 & & 1127.098 & 0.483$^{{\rm f}}$ & -41.8 $\pm$ 1.5 & 23.7 $\pm$ 2.8 & 
 14.97 $\pm$ 0.06 \\
 & & 1611.201 & 0.339 & -42.5 $\pm$ 1.3 & 24.8 $\pm$ 3.3 & 
 14.98 $\pm$ 0.06 \\
 & & & & & &  \\
Al {\small III}\tablenotemark{g} & 18.8$-$28.5 & 1854.716 & 3.017 & -50.1 $\pm$ 4.2 & 
166 $\pm$ 17 & 13.05 $\pm$ 0.05 \\
 & & 1862.790 & 2.716 & -68.5 $\pm$ 6.4 & 76 $\pm$ 15 & 13.01 $\pm$ 0.10  \\
 & & & & & & \\
Fe {\small III} & 16.2$-$30.7 & 1122.546 & 2.260 & -61.7 $\pm$ 0.9 & 58.1 $\pm$ 3.9 & 
13.61 $\pm$ 0.04 \\
 & & & & & &  \\
Si {\small IV} & 33.5$-$45.1 & 1393.755 & 2.855 & -66.6 $\pm$ 0.4 & 110.3 $\pm$ 3.0 & 
13.32 $\pm$ 0.02 \\
 & & 1402.770 & 2.554 & -67.5 $\pm$ 0.5 & 72.5 $\pm$ 2.3 &  13.35 $\pm$ 0.02 \\
 & & & & & &  \\
C {\small IV} & 47.9$-$64.5 & 1548.195 & 2.470 & -65.7 $\pm$ 0.5 & 151.0 $\pm$ 4.8  &  
13.89 $\pm$ 0.04  \\
 & & 1550.770 & 2.169 & -66.5 $\pm$ 0.5 & 114.8 $\pm$ 3.8 & 13.93 $\pm$ 0.02 \\
 & & & & & & \\
N {\small V} & 77.5$-$97.9 & 1238.821 & 2.289 & $\ldots$ & $\leq$ 
10.6  &  $\leq$ 12.72$^{{\rm h}}$ \\ 
 & & 1242.804 & 1.988 & $\ldots$ & $\leq$ 4.7 &  $\leq$ 12.65$^{{\rm h}}$  \\ 
 & & & & & &  \\
O {\small VI} & 113.9$-$138.1 & 1031.926 & 2.137 & $\ldots$ & $\leq$ 
54.4$^{{\rm i}}$ & $\leq$ 13.78$^{{\rm h}}$ \\ 
 & & & & & &  \\ 
\enddata
\footnotesize{
\tablenotetext{a}{Creation and destruction ionization potentials (Lang 1980).}
\tablenotetext{b}{Morton (1991).}
\tablenotetext{c}{1-$\sigma$ $W_{\lambda}$ uncertainties and upper limits 
derived following the prescription in Morton, York, \& Jenkins (1986).}
\tablenotetext{d}{$N$ derived from the apparent optical depth method
(Sembach \& Savage 1992).} 
\tablenotetext{e}{Oscillator strengths from Sofia, Fabian, \& Howk (2000).}
\tablenotetext{f}{Howk et al. (2000).}
\tablenotetext{g}{Savage, Meade, \& Sembach (2001) -- based on
high-resolution (R $\sim$ 25 km s$^{-1}$) $IUE$ observations.}
\tablenotetext{h}{3-$\sigma$ upper limit assuming a $b$-value of 
12.3 km s$^{-1}$ derived from \ion{C}{4} and \ion{Si}{4} results.}
\tablenotetext{i}{\ion{O}{6} upper limit includes absorption by a possible
stellar wind feature (see text).}} 
\end{deluxetable}

\newpage
\clearpage
\setcounter{table}{2}
\begin{deluxetable}{ccccccc}
\tablecolumns{6}
\tablewidth{0pt}
\tablecaption{Profile Synthesis Results for \ion{Si}{4} and \ion{C}{4} 
                \label{CIV_SiIVcomponents}}
\tablehead{ 
\colhead{$V_{LSR}$} & \colhead{log[$N$(Si {\small IV})]} & 
\colhead{log[$N$(C {\small IV})]} & 
\colhead{{\Large $\frac { N({\rm C~IV}) } { N({\rm Si~IV}) } $ } } & 
\colhead{v$_{{\rm nt}}$\tablenotemark{{\rm a}}} & \colhead{$T$} & $\chi^2$/$\nu$ \\
\colhead{[km s$^{-1}$]} & \colhead{[cm$^{-2}$]} & 
\colhead{[cm$^{-2}$]} & \colhead{} & 
\colhead{[km s$^{-1}$]} & \colhead{[K]} & }
\startdata
\cutinhead{One Component Model}
-67.3 $\pm$ 0.1$^{\rm b}$ & 13.35 $\pm$ 0.01 &
13.93 $\pm$ 0.01 & 3.8 $\pm$ 0.1 &
12.0 $\pm$ 0.2 & 10,450 $\pm$ 3,400 & 1.42 \\
 & & & & & & \\ 
\cutinhead{Two Component Model\tablenotemark{{\rm b}}} 
-65.8 & 13.16 & 13.81 & 4.4 & 10.7 & 26,500 & 1.43 \\ 
-71.2 & 12.89 & 13.31 & 2.6 & 11.3 & 10,100 & \\
 & & & & & & \\
\enddata
\tablenotetext{a}{Non thermal velocity.} 
\tablenotetext{b}{1-$\sigma$ errors do not include continuum placement
uncertainties.  No errors are reported for the two component model due to
the low confidence in the second component.} 
\end{deluxetable}

\setcounter{table}{3}
\begin{deluxetable}{cccc}
\tablecolumns{4}
\tablewidth{0pt}
\tablecaption{Adopted Parameters of IVC Absorption \label{ISM_den}}
\tablehead{ 
\colhead{Species} & \colhead{$b$-value} & \colhead{$V_{LSR}$} &
\colhead{log($N$)\tablenotemark{a}} \\
\colhead{} & \colhead{[km s$^{-1}$]} & \colhead{[km s$^{-1}$]} &
\colhead{[cm$^{-2}$]} }
\startdata
Mg {\small II} & $\ldots$ & -47.1 $\pm$ 1.1 & 15.75 $\pm$ 0.04 \\
Fe {\small II} &  23.0 $\pm$ 4.4 & -46.2 $\pm$ 0.9 & 14.90 $\pm$ 0.06\tablenotemark{b} \\
Al {\small III}\tablenotemark{c}$^{~,}$\tablenotemark{d} & $\ldots$ & -56.3 $\pm$ 3.3 & 13.04 $\pm$ 0.05 \\
Fe {\small III}\tablenotemark{d} &  $\ldots$ & -61.7 $\pm$ 0.9 & 13.61 $\pm$ 0.04 \\
Si {\small IV} & 12.3 $\pm$ 0.3 & -66.6 $\pm$ 0.4 & 13.34 $\pm$ 0.02 \\
C {\small IV} & 12.6 $\pm$ 0.3  & -65.7 $\pm$ 0.5 & 13.92 $\pm$ 0.02 \\
N {\small V} & $\ldots$ & $\ldots$ &  $\leq$ 12.65\tablenotemark{e} \\ 
O {\small VI} & $\ldots$ & $\ldots$ & $\leq$ 13.73\tablenotemark{e}$^{~,}$\tablenotemark{f} \\ 
 & & &  \\ 
\enddata
\footnotesize{
\tablenotetext{a}{Weighted average of $N_a$($v$) from Table~\ref{ISM_par}.}
\tablenotetext{b}{Derived from curve of growth (see text for details).}
\tablenotetext{c}{Savage, Meade, \& Sembach (2001).}
\tablenotetext{d}{Contains contribution from both IVC and Perseus Arm material.}
\tablenotetext{e}{Upper limit to $N$ determined with an assumed $b$-value of 
12.3 km s$^{-1}$ derived from average \ion{C}{4} and \ion{Si}{4} results.}
\tablenotetext{f}{\ion{O}{6} upper limit includes effect of possible
stellar wind feature (see text).}} 
\end{deluxetable}

\begin{deluxetable}{ccccccc}
\setcounter{table}{4}
\tablecolumns{6}
\tablewidth{0pt}
\tablecaption{\ion{H}{2} Region Photoionization 
  Models\tablenotemark{a}\label{CLOUDY}}
\tablehead{ 
\colhead{Model} & 
\colhead{$\log n_{\rm H}$} & 
\colhead{$\log U$\tablenotemark{b}} & \colhead{log[$N$(C~IV)]} & 
\colhead{{\Large $\frac { N({\rm C~IV}) } { N({\rm Si~IV}) } $ } } &
\colhead{{\Large $\frac { N({\rm Al~III}) } { N({\rm Fe~III}) } $ } } \\
\colhead{} & \colhead{[cm$^{-3}$]} & \colhead{} & \colhead{[cm$^{-2}$]} &  \colhead{} & \colhead{} 
}
\startdata
 1  &  -4.0  &   1.3  & 8.0 & 0.79  &  0.05   \\
 2  &  -2.0  &  -0.7  & 8.6 & 0.77  &  0.06   \\
 3  &   0.0  &  -2.7  & 10.3 & 0.43  &  0.04   \\
\enddata
\tablenotetext{a}{All column densities assume abundances appropriate
  for the WNM following Sembach et al. (2000).}
\tablenotetext{b}{The ionization parameter: the dimensionless 
  ratio of photon to particle densities.}
\end{deluxetable}

\setcounter{table}{5}
\begin{deluxetable}{lccccc}
\tablecolumns{6}
\tablewidth{0pt}
\tablecaption{Hot Plasma Models\label{localbubble}}
\tablehead{\colhead{Quantity} & \colhead{Observed} & 
\colhead{1} & \colhead{2} & 
\colhead{3\tablenotemark{b}} & 
\colhead{4}} 
\startdata
Abundances\tablenotemark{a} & \nodata & Solar & WNM & WNM & WNM \\
\\
$\log(T_{\rm plasma})$ [K] & \nodata & 6.1 & 6.1 & 6.1 & 7.0 \\
\\
$\log(n_{\rm H})$ [cm$^{-3}$] & \nodata & -2.9 & -2.8 & -1.8 & -0.7 \\
\\
$\log(U)$ & $\ldots$ & -1.6 & -1.7 & -2.9 & -1.7 \\
\\
$\log[N({\rm H}_{tot})]$ [cm$^{-2}$] & \nodata & 18.55 & 19.01 & 18.92 & 
18.99 \\
\\
$T$ [K] & 10,450 $\pm$ 3,400 & 4,630 & 6,930 & 10,100 & 6,880 \\
\\
log[$N$({\rm C~IV})] [cm$^{-2}$] &  13.92 $\pm$ 0.02 & 13.92 & 13.92 & 13.92 &
13.92 \\
\\
{\large $\frac { N({\rm C~IV}) } { N({\rm Si~IV}) } $ } & 3.8 $\pm$ 0.3 & 3.8 & 
3.8 & 3.7 & 3.8 \\
\\
{\large $\frac { N({\rm C~IV}) } { N({\rm N~V}) } $ } & $\geq$ 18.6 & 194 & 180
 & 89 & 171\\
\\
{\large $\frac { N({\rm C~IV}) } { N({\rm O~VI}) } $ } & $\geq$ 1.6 & 6341 & 
9749 & 1475 & 8165\\
\\
{\large $\frac { N({\rm C~IV}) } { N({\rm Al~III}) } $ } & $\geq$ 7.6 & 18 & 
126 & 331 & 138\\
\\
{\large $\frac { N({\rm C~IV}) } { N({\rm Fe~III}) } $ } & $\geq$ 2.0 & 25 & 
139 & 49 & 163\\
\\
{\large $\frac { N({\rm Al~III}) } { N({\rm Fe~III}) } $ } & 0.3 $\pm$
0.1\tablenotemark{c}& 1.4 & 1.1 & 0.2 & 1.2 \\
\enddata
\tablenotetext{a}{Assumed to be Solar or depleted abundances found for the
	WNM (see text).  Solar abundances use Anders \& Grevesse (1989) summary
	with the oxygen abundance of Holweger (2001) and do not include 
	interstellar grains.  The WNM abundances adopt the relative abundances 
	summarized in Sembach et al. (2000) and includes interstellar grains.}
\tablenotetext{b}{Uses X-ray flux a factor-of-10 higher that observed in the 
	Local Bubble used to compute models 1 and 2
	(see Figure~\ref{ion_input}).}
\tablenotetext{c}{Contains contribution from both IVC and Perseus Arm gas 
	because both components are blended together.}
\end{deluxetable}
\clearpage

\end{document}